\documentclass[aps,prl,twocolumn,superscriptaddress]{revtex4}

\usepackage{epsf,epsfig,amsmath,amssymb}

\newsavebox{\ns}
\newsavebox{\dbrane}

\def\be{\begin{equation}}
\def\ee{\end{equation}}
\def\bea{\begin{eqnarray}}
\def\eea{\end{eqnarray}}

\def\Dslash{\,\,{\raise.15ex\hbox{/}\mkern-12mu D}}
\def\Dbarslash{\,\,{\raise.15ex\hbox{/}\mkern-12mu {\bar D}}}
\def\delslash{\,\,{\raise.15ex\hbox{/}\mkern-9mu \partial}}
\def\delbarslash{\,\,{\raise.15ex\hbox{/}\mkern-9mu {\bar\partial}}}
\def\pslash{\,\,{\raise.15ex\hbox{/}\mkern-9mu p}}
\def\calDslash{\,\,{\raise.15ex\hbox{/}\mkern-12mu {\cal D}}}

\newcommand{\vol}{\mbox{vol}}

\newcommand{\nn}{\nonumber \\}
\newcommand{\reef}[1]{(\ref{#1})}

\newcommand{\bbR}{\mathbb{R}}

\begin{document}

\title{Holographic superconductivity in M-Theory}

\author{Jerome P. Gauntlett, Julian Sonner and Toby Wiseman}
\affiliation{Theoretical Physics Group, Blackett Laboratory,
  Imperial College, London SW7 2AZ, U.K.}
\affiliation{The Institute for Mathematical Sciences,
  Imperial College, London SW7 2PE, U.K.}

\begin{abstract}
Using seven-dimensional Sasaki-Einstein spaces we construct solutions of $D=11$ supergravity 
that are holographically dual to superconductors in three spacetime dimensions.
Our numerical results indicate a new zero temperature solution dual to a quantum critical point.
\end{abstract} 

\maketitle

\section{Introduction}
\vspace{-0.3cm}
The AdS/CFT correspondence provides a powerful framework
for studying strongly coupled quantum field theories using
gravitational techniques. It is an exciting possibility that
these techniques can be used to study classes of superconductors which are not
well described by more standard approaches \cite{Gubser:2008px}\cite{H31}\cite{H32}.

The basic setup requires that the CFT has a
global abelian symmetry corresponding to a massless gauge field propagating in the
$AdS$ space. We also require an operator in the CFT that corresponds to a scalar field that
is charged with respect to this gauge field. Adding a black hole to the $AdS$ space describes the
CFT at finite temperature. One then looks for cases where there are high temperature black hole
solutions with no charged scalar hair but below some critical temperature black hole solutions
with charged scalar hair appear and moreover dominate the free energy.
Since we are interested in describing superconductors in flat spacetime we consider black holes with planar symmetry.
In order to obtain a critical temperature, conformal
invariance then implies that another scale needs to be introduced. 
This is achieved by considering electrically charged black holes 
which corresponds to studying the dual CFT at finite chemical potential.

Precisely this set up has been studied 
using a phenomenological theory of gravity
in $D=4$ coupled to a single charged scalar field and it has been shown that, for certain parameters,
the system manifests superconductivity in three spacetime dimensions, in the above sense
\cite{H32}. It is important to go beyond such models and construct solutions in the context of 
string/M-theory so that there is a consistent underlying quantum theory 
and CFT dual.
Also, as we shall see, the behaviour of the string/M-theory solutions will differ substantially from that of the phenomenological model \cite{H32} at low temperature.
It was shown in \cite{dh} that the $D=4$ phenomenological models 
of \cite{H32} arise, at the linearised level, after Kaluza-Klein (KK) reduction of $D=11$ supergravity 
on a seven-dimensional Sasaki-Einstein space $SE_7$.  
Here we go beyond this linearised approximation by working with a consistent truncation of the $D=4$ KK reduced theory
presented in \cite{Gauntlett:2009zw}. The truncation is consistent in the sense that any solution of this
$D=4$ theory, combined with a given $SE_7$ metric, gives rise to an exact solution of $D=11$ supergravity.
Here we shall use this $D=4$ theory to construct exact solutions of $D=11$ supergravity that correspond 
to holographic superconductivity. 

\vspace{-0.5cm}
\section{The KK truncation}
\vspace{-0.3cm}

We begin by recalling that any $SE_7$ metric
can, locally, be written as a fibration over a six-dimensional K\"ahler-Einstein space, $KE_6$:
\be
ds^2(SE_7)\equiv ds^2(KE_6)+\eta\otimes\eta
\ee
Here $\eta$ is the one-form dual to the Reeb Killing vector satisfying
$d\eta=2J$ where $J$ is the K\"ahler form of $KE_6$. We denote the $(3,0)$ form defined
on $KE_6$ by $\Omega$. For a regular or quasi-regular $SE_7$ 
manifold, the orbits of the
Reeb vector all close, corresponding to compact $U(1)$ isometry,
and the $KE_6$ is a globally defined manifold or
orbifold, respectively. For an irregular $SE_7$ manifold, the Reeb-vector generates a
non-compact $\bbR$ isometry and the $KE_6$ is only locally defined.

In the KK ansatz of \cite{Gauntlett:2009zw} the $D=11$ metric is written
\be\label{KKmetT}
ds^2=e^{-6U-V}ds^2_4+e^{2U}ds^2(KE_6)+e^{2V}(\eta+A_1)\otimes(\eta +A_1) 
\ee
while the four-form is written
\bea
\label{KKG4T}
   G_4 &= 6 e^{-18U-3V}\left(
           \epsilon+h^2+|\chi|^2\right)\vol_4 + H_3 \wedge(\eta+A_1)\nn
      &+ H_2 \wedge J
             + dh \wedge J \wedge (\eta+A_1) + 2h J \wedge J\nn
             &+ {\sqrt 3}\left[
         \chi(\eta+A_1)\wedge\Omega-\tfrac{i}{4}D\chi\wedge\Omega
         + \textrm{c.c.} \right] 
\eea
where $ds^2_4$ is a four-dimensional metric (in Einstein frame),
$U,V,h$ are real scalars, and $\chi$ is a
complex scalar defined on the four-dimensional space.
Furthermore, also defined on this four-dimensional space are $A_1$ a one-form potential, with field strength $F_2\equiv dA_1$,
two-form and three-form field strengths $H_2$ and $H_3$, related to one-form and two-form potentials
via $H_3=dB_2$ and $H_2= dB_1+2B_2+hF_2$.
Finally $D\chi\equiv d\chi-4iA_1\chi$.

This is a consistent KK truncation of $D=11$ supergravity in the sense that if the
equations of motion for the 4d-fields $ds^2_4,U,V,A_1,H_2,H_3,h,\chi$
as given in \cite{Gauntlett:2009zw} are satisfied then so are 
the $D=11$ equations.
The $D=4$ equations of motion admit a vacuum solution 
with vanishing matter fields
which uplifts to the
$D=11$ solution:
\bea\label{d11vac}
ds^2=\tfrac{1}{4}ds^2(AdS_4)+ds^2(SE_7),\quad
G_4=\epsilon\tfrac{3}{8}Vol(AdS_4)
\eea
where $ds^2(AdS)_4$ is the standard unit radius metric.
When $\epsilon=+1$, this $AdS_4\times SE_7$ solution is supersymmetric and
describes $M2$-branes sitting at the apex of the Calabi-Yau four-fold ($CY_4$)
cone whose base space is given by the $SE_7$.
When $\epsilon=-1$ the solution is a ``skew-whiffed'' $AdS_4\times SE_7$ solution, 
which describes anti-$M2$-branes sitting at the apex of the $CY_4$ cone. 
These solutions break all of the supersymmetry
except for the special case when the $SE_7$ is the round seven-sphere, $S^7$, 
in which case it is maximally supersymmetric.
Note that the skew-whiffed solutions with $SE_7\ne S^7$ are 
perturbatively stable \cite{Duff:1984sv}, despite the absence of 
supersymmetry. Thus such backgrounds should be dual to 
three-dimensional CFTs at least in the strict $N=\infty$ limit.
We are most interested in the skew-whiffed case because
it is for that case that the operator dual to $\chi$ has 
scaling dimensions $\Delta=1$ or $2$ 
\cite{Gauntlett:2009zw} and, based on the work of \cite{dh}, is when we expect holographic superconductivity.

The $D=4$ equations of motion can be derived from a four-dimensional action given in \cite{Gauntlett:2009zw}.
It is convenient to work with an action that is obtained after dualising
the one-form $B_1$ to another one form $\tilde B_1$ and the two-form $B_2$ to a scalar $a$
as explained in section 2.3 of \cite{Gauntlett:2009zw}. The dual fields are related to the original fields via
\bea
H_3&=&-e^{-12U}*\left[da+6(\tilde B_1-\epsilon A_1)-\tfrac{3}{4}i(\chi^* D\chi-\chi D\chi^*)\right]\nn
H_2&=&(4h^2+e^{4U+2V})^{-1}\left[2h-e^{2U+V}*\right](\tilde H_2+h^2 F_2)
\eea
where $\tilde H_2\equiv d\tilde B_1$.
We now restrict to the (skew-whiffed) case $\epsilon=-1$. For this case
we can make the following additional truncation of the $D=4$ theory:
\bea\label{truncansatz}
a&=&h=0,\qquad V=-2U,\nn A_1 &=& -\tilde B_1,\qquad e^{6U}=1-\tfrac{3}{4}|\chi|^2
\eea
One can show that provided that we restrict to configurations
satisfying ${F_2}\wedge {F_2}=0$ we obtain equations of motion 
that can be derived from the action
\bea
\label{lageinfull}
  S = \frac{1}{16\pi G}\int d^4 x\sqrt{-g}\Big[
      R - \tfrac{1}{4}\hat{F}_{\mu\nu} 
\hat{F}^{\mu\nu} \qquad \qquad \qquad \nn
+ (1-\tfrac{1}{2}|\hat\chi|^2)^{-2}\left( - |D\hat{\chi}|^2 
+ 24 (1-\tfrac{2}{3}|\hat{\chi}|^2) \right)
      \Big]
\eea
where $D\hat\chi\equiv d\hat\chi-2i\hat A_1 \hat\chi$,
and we have defined $\hat{A_1}\equiv 2A_1$,
$\hat{\chi}\equiv (3/2)^{1/2}\chi$.
Linearizing in the complex scalar $\hat\chi$, this gives the action considered in
\cite{H32} (with their $L = 1/2$ and their $q=2$). 
This non-linear action is in the class considered in \cite{Franco:2009yz} and
in addition to the $AdS_4$ vacuum with $\hat A_1=0$ and $\hat \chi=0$, which uplifts
to \reef{d11vac}, it also admits $AdS_4$ vacuua
with $\hat A_1=0$ and constant $|\hat\chi|=1$, 
which uplift to the 
$D=11$ solutions\footnote{There are analogous $AdS_5$ solutions 
of the theory considered in 
\cite{Gubser:2009qm} which uplift to IIB solutions found in \cite{romans}.}
found in \cite{pw}.

\vspace{-0.5cm}
\section{Black Hole Solutions}
\vspace{-0.3cm}

The key result of the last section is that any solution to the 
$D=4$ equations of motion of the action \reef{lageinfull} with $\hat{F}\wedge \hat{F}=0$,
gives an exact solution of $D=11$ supergravity for any $SE_7$ metric.
To find solutions relevant for studying superconductivity via holography 
we consider the following ansatz
\newcommand\f{{\hat{\phi}}}
\newcommand\G{{g}}
\newcommand\B{{\beta}}
\newcommand\hmu{ {\hat{\mu}}  } 
\newcommand\hq{{\hat{q}}    }
\bea
ds^2&=&-g e^{-\B}dt^2+g^{-1}dr^2+r^2(dx^2+dy^2)\nn
\hat{A}_1&=&\f dt,\qquad \hat{\chi}\equiv\sigma \in\bbR
\eea
where $g,\beta,\f$ and $\sigma$ are all functions of $r$ only. Being purely electrically charged this satisfies the $\hat{F} \wedge \hat{F} = 0$ condition.
After substituting into the equations of motion arising from
\reef{lageinfull}, we are led to
ordinary differential equations which can also be obtained from the action 
obtained by substituting the ansatz directly into 
\reef{lageinfull}:
\bea
S&=&c\int dr r^2e^{-\beta/2}\Big[-g''+g'(\frac{3}{2}\beta'-\frac{4}{r}) \nn
&+&g(\beta''-\frac{1}{2}(\beta')^2+2\frac{\beta'}{r}-\frac{2}{r^2})
+\frac{1}{2}e^{\beta}(\f')^2 \\
& + & (1-\tfrac{1}{2}\sigma^2)^{-2} \left( - g(\sigma')^2
+ 4g^{-1}e^{\beta}\f^2\sigma^2
+24(1-\tfrac{2}{3}\sigma^2) \right) \Big] \nonumber
\eea
where $c=(16\pi G)^{-1}\int dt dx dy$.

We next observe that the system admits the following exact
AdS Reissner-Nordstr\"om type solution $\sigma=\beta=0$
\bea\label{asoln}
g=4r^2-\frac{1}{r}(4r_+^3+\frac{\alpha^2}{r_+})+\frac{\alpha^2}{r^2},\quad
\f=\alpha(\frac{1}{r_+}-\frac{1}{r})
\eea
for some constants $\alpha,r_+$. The horizon is located at $r=r_+$ and for large $r$
it asymptotically approaches 1/4 of a unit radius $AdS_4$ (see \reef{d11vac}).
This solution should describe the high temperature phase of the superconductor.

We are interested in finding more general black hole solutions with charged scalar hair, $\sigma\ne 0$.
Let us examine the equations at the horizon and at infinity.
At the horizon $r=r_+$ we demand that
$g(r_+)=\f(r_+)=0$.
One then finds that the solution is specified by 4 parameters at the horizon
$r_+$, $\beta(r_+)$, $\f'(r_+)$, $\sigma(r_+)$.
At $r=\infty$ we have the asymptotic expansion,
\begin{eqnarray}
 \beta =  \beta_a + \ldots,
\; \frac{\sigma}{\sqrt{8\pi G}}  =  \frac{ \sigma_1 }{r} + 
\frac{ \sigma_2}{r^2} + \ldots , \nn \frac{\f}{\sqrt{16\pi G}} = e^{-\beta_a/2} 
( \hmu - \frac{\hq}{r}) + \ldots \nonumber 
 \end{eqnarray}
 \vspace{-0.5cm}
\begin{eqnarray}\label{aexpc}
e^{-\beta} g = e^{-\beta_a} ( 4 r^2  - \frac{8\pi G(m+\frac{4}{3}\sigma_1\sigma_2)}{r} ) + \ldots \qquad \qquad  \qquad  \qquad  \qquad
\end{eqnarray}
determined by the data $\beta_a, \sigma_{1,2}, m, \hat{\mu}, \hat{q}$.
The scaling
\be\label{sc1}
r\to ar,\,(t,x,y)\to a^{-1}(t,x,y),\, g\to a^2g,\,
\f\to a\f
\ee
leaves the metric, $A_1$, and all equations of motion
invariant.
 
\vspace{-0.7cm}
\subsection{Action and thermodynamics}
\vspace{-0.3cm}

We analytically continue by defining $\tau\equiv i t$. The temperature of the black hole is
$T = e^{\beta_a/2}/\Delta \tau$ where $\Delta\tau$ is fixed by demanding regularity 
of the Euclidean metric at $r=r_+$. We find:
\bea
T=
\frac{r_+e^{(\beta_a-\beta)/2}}{4\pi}
\Big[  \frac{12 \, (1-\tfrac{2}{3}\sigma^2)}{(1-\tfrac{1}{2}\sigma^2)^2}  - \tfrac{1}{4} e^{\beta} \f'^2 \Big]_{r=r_+}
\eea
Defining $I\equiv -iS$, we can calculate the on-shell
Euclidean action $I_{OS}$
\bea\label{actone}
I_{OS}
&=&\frac{\Delta \tau vol_2}{16\pi G}\int_{r_+}^{\infty}dr \left[r^2e^{-\beta/2}(g'-g\beta'-e^{\beta}\f\f'
)\right]'\nn
&=&\frac{\Delta \tau vol_2}{16\pi G}\int_{r_+}^{\infty}dr \left[2r\G e^{-\B/2}\right]'
\eea
where $vol_2\equiv \int dx dy$. The latter expression only gets 
contributions from the 
on-shell functions at $r=\infty$ since $g(r_+)=0$, while the 
former expression gets contributions from $r=r_+$ and $r=\infty$.
\newcommand\ce{{\frac{\Delta \tau vol_2}{16\pi G}}}
The on-shell action diverges and we need to regulate by adding appropriate counter terms.
We define $I_{Tot}\equiv I+I_{ct}$ and, for simplicity, we will focus on the following counter-term action $I_{ct}$:
\be
I_{ct}=\frac{1}{16\pi G}\int d\tau d^2x {\sqrt{g_\infty}}\left[-2K+8+2\sigma^2\right]
\ee
where $K=g^{\hmu\nu}_\infty\nabla_\hmu n_\nu$ is the trace of the extrinsic curvature.
For our class of solutions we find
\begin{eqnarray}
I_{ct}&=&\frac{\Delta \tau vol_2}{16\pi G}
\lim_{r \rightarrow \infty}e^{-\beta/2}\Big[-r^2g' +r^2g\beta'-4gr \nn
&& \qquad \qquad \qquad  \qquad  \qquad +r^2g^{1/2}(8+2\sigma^2)\Big]
\end{eqnarray}
Notice that under a variation of the action $I_{Tot}$ with respect to $\beta,g,\f$
yields the equations of motion together with surface terms.
For an on-shell variation the only terms remaining are these surface terms, 
and after substituting the asympototic boundary expansion \reef{aexpc}
(higher order terms are also required) we find
\begin{eqnarray}\label{osvary}
[ \delta I_{Tot} ]_{OS}  & = & \frac{\Delta \tau vol_2}{16 \pi} e^{-\beta_a/2} \Big[  
( - \tfrac{1}{2}m + \tfrac{1}{2} \hmu \, \hq ) \delta \beta_a \nn
&& \qquad \qquad  - \hq \, \delta \hmu - 4\sigma_2  \delta \sigma_1 
 \Big] 
\end{eqnarray}
Note that we are keeping $\Delta\tau$ fixed in this variation.
Hence we see that $I_{Tot}$ 
is stationary for fixed temperature and chemical potential (ie. $\delta \beta_a = \delta \hmu = 0$) and for either $\sigma_2 = 0$ or fixed $\sigma_1$.

We also find that the on-shell total action is given by
\begin{eqnarray}\label{osact}
[ I_{Tot} ]_{OS} & = & \frac{vol_2}{T}  \Big[ m 
- \, \hmu \, \hq - Ts \Big]   \nn
& = & 
\frac{vol_2}{T} \Big[ - \frac{1}{2} m - 2\sigma_1 \sigma_2 \Big]
\end{eqnarray}
where $s = \frac{r_+^2}{4 G}$ is the entropy density
of the solution and $m$ is the energy density. 
The two forms of the on-shell action come from the two ways of writing the action as a total derivative given above. 
We note that the equality of these expressions imply a Smarr-like relation.
Also note that after using $\delta \beta_a = 2 \, \delta T / T$ (since $\Delta\tau$ is held fixed)
the equality of \eqref{osvary} and the variation of the first line of  \eqref{osact} imply 
a first law,
\begin{eqnarray}
\delta m = T \delta s +  \, \hmu \, \delta \hq - 4\sigma_2 \delta \sigma_1 \, .
\end{eqnarray}
Both this Smarr relation and the first law were used to confirm the 
accuracy of our numerical solutions below.

For simplicity we restrict discussion to solutions with boundary condition $\sigma_1=0$ \footnote{One may also consider fixing $\sigma_2 = 0$ \cite{H32}
with similar results. Non-zero $\sigma_1$ is less interesting as we want the scalar to condense without being sourced.} 
and we interpret $TI_{Tot}=(vol_2)(-m/2)$ 
as a thermodynamic potential, $\Omega(T,\mu)$.
Note also that $\sigma_2$ then determines the vev of the operator dual to $\chi$. 
Recall from \cite{Gauntlett:2009zw} that writing $U=-u+v/3$, $V=6u+v/3$, the
fields $u,v$ are dual to operators $\mathcal{O}_{u,v}$ with dimensions $\Delta_u=4$, $\Delta_v=6$.
The truncation \reef{truncansatz} implies that the vevs of these
dual operators are fixed by $\sigma_2$. The asympotic expansion of
$u$ to $o(1/r^4)$ and $v$ to $o(1/r^6)$ 
gives $\langle \, \mathcal{O}_{u} \, \rangle \propto \sigma_2^2$ 
and $ \langle \mathcal{O}_{v} \, \rangle\propto  \mu^2 \sigma_2 ^2 $.

\vspace{-0.5cm}
\subsection{Numerical Results}
\vspace{-0.3cm}

Following \cite{H32} we 
solved the differential equations numerically using a shooting method. 
We used \reef{sc1} to fix the scale $\hat{\mu}=1$.
At high temperatures the black hole solutions have no scalar hair ($\sigma_2=0$) and are just the
solutions given in \reef{asoln}. 
At a critical temperature $T_c\sim 0.042$ a new branch of solutions
with $\sigma_2\ne 0$ appears and moreover dominates the free energy. We refer to these as the unbroken and
broken phase solutions, corresponding to normal and superconducting phases, respectively.
In the figures we have plotted some features of our solutions and compared them with the solutions of the phenomenological model
considered in \cite{H32}. 

While the results are in agreement near the critical temperature,
as expected, we see marked differences as the temperature goes to zero. 
We have calculated
the Ricci scalar and curvature invariant $\sqrt{R_{\alpha\beta\mu\nu} R^{\alpha\beta\mu\nu}}$ at $r=r_+$
which indicate that the solutions of \cite{H32} are becoming singular
but our solutions are approaching a 
regular zero temperature solution, without horizon,
holographically dual to a quantum critical point. 
Indeed as $r\to r_+$ we find 
$\sigma \sim 1$, $\beta \sim $const, $\hat\phi \sim 0$ and $g \sim \tfrac{16}{3} \left( r^2 - r_+^3 / r \right)$ and fixing $\hat{\mu} = 1$ gives $r_+ \rightarrow 0$ in the extremal limit. 
In particular, the geometry near $r=r_+$ is consistent with being the exact $AdS_4$ 
solution with $\sigma= 1$, mentioned earlier, which uplifts to the $D=11$ solution found in \cite{pw}.
For such a solution $R = -64$ and $\sqrt{R_{\alpha\beta\mu\nu} R^{\alpha\beta\mu\nu}} = 32$, agreeing with the low temperature limit seen in the figures. 
The full zero temperature solution thus appears to be a charged
domain wall, of the type considered in \cite{Gubser:2008wz},
connecting two $AdS_4$ vacua of \eqref{lageinfull}, one 
with $\sigma=0$ and the other with $\sigma=1$. 
Interestingly this implies the entropy of the solutions vanish in the low temperature limit, unlike for the 
Reissner-Nordstr\"om solution \eqref{asoln}.\footnote{We thank Gary Horowitz for a discussion on this point.} 
The asymptotic charge appears to be derived from the scalar hair, with 
the region near $r=r_+$ carrying no flux.

\begin{figure}
\includegraphics[height=2.2in,width=3.0in]{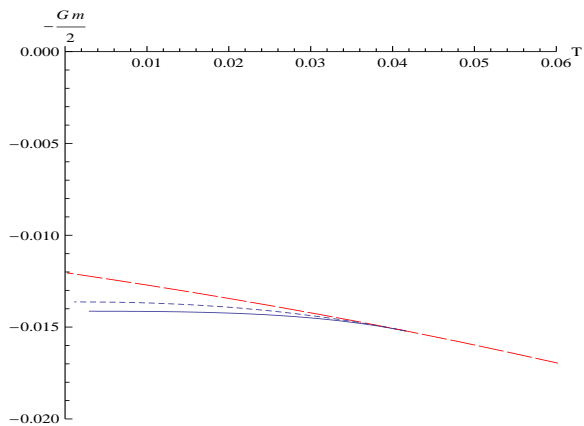}
\caption{
Plot showing $-\tfrac{1}{2} Gm$ (proportional to the 
thermodynamic potential $\Omega(T,\mu)$) against $T$ with fixed $\hat{\mu} = 1$, for
unbroken phase solutions (long dashed red), broken phase (blue) and solutions of \cite{H32} (with their $L=1/2$ and their $q=2$) (dashed blue). 
}
\end{figure}
\begin{figure}
\includegraphics[height=2.4in,width=3.0in]{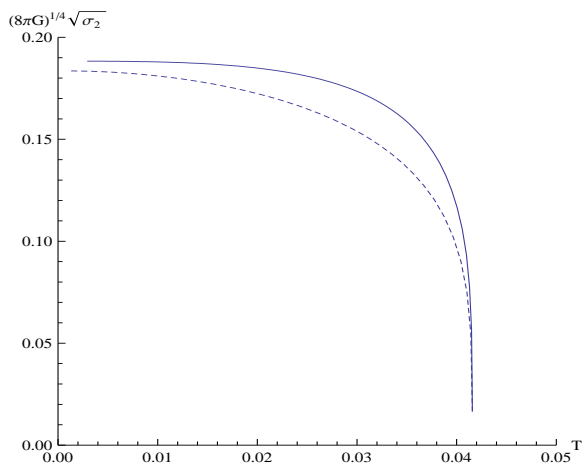}
\caption{
Plot showing the asymptotic value of the scalar condensate, 
$(8\pi G)^{1/4}\sqrt{\sigma_2}$, against $T$ (conventions as above).
}
\end{figure}
\begin{figure}
\includegraphics[height=2.4in,width=3.0in]{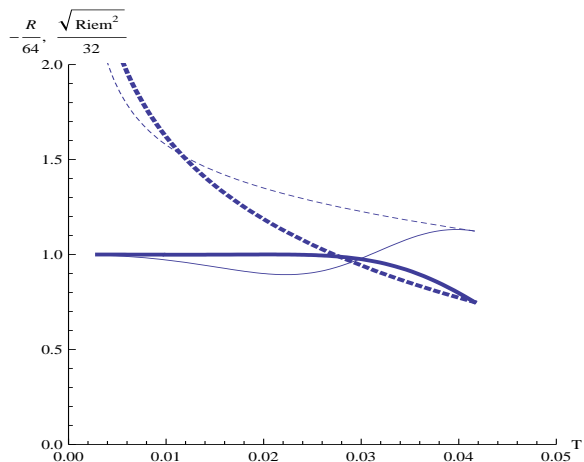}
\caption{
Plots showing the value of the Ricci scalar (heavy lines) and $\sqrt{R_{\alpha\beta\mu\nu}R^{\alpha\beta\mu\nu}}$ (light lines) at the horizon normalized by $-64$ and $32$ respectively. (conventions as above).
}
\end{figure}

\vspace{-0.5cm}
\section{Concluding remarks}
\vspace{-0.3cm}

For any seven-dimensional Sasaki-Einstein space we have constructed solutions of $D=11$ supergravity corresponding to holographic 
superconductors in three spacetime dimensions. 
We have studied electric 
black holes using the action \eqref{lageinfull} whose 
solutions lift to $D=11$ when $\hat{F} \wedge \hat{F} = 0$. One may consider adding magnetic charge using the full consistent truncation of \cite{Gauntlett:2009zw}.
Our results indicate the existence of a regular zero temperature solution which is a charged domain wall connecting two $AdS_4$ vacua of \eqref{lageinfull} and dual to
a new quantum critical point.
An important open issue is whether or not there are additional unstable charged modes for skew-whiffed $AdS_4\times SE_7$ solutions, which condense
at higher temperatures. If they exist, and dominate the free energy, then the corresponding supergravity solutions would be the 
appropriate ones to describe the superconductivity and not the ones that we have constructed. However, 
it is plausible that we have found the dominant modes for large classes of $SE_7$, if not all.
For the specific class of deformations of the four-form that were considered in \cite{dh}, it was proven that the modes that we consider
are in fact the only condensing modes. It would be worthwhile extending this result to cover other bosonic and/or fermionic modes.

{\it Note added:} after this work was completed we received  \cite{Gubser:2009qm} which constructs solutions of string theory that are dual to superconductors in four spacetime dimensions.


\medskip
\noindent
We are supported by EPSRC (JG,JS), the Royal Society (JG) and STFC (TW).
We thank R. Emparan, S. Hartnoll, G. Horowitz, V. Hubeny, M. Rangamani, 
O. Varela and D. Waldram for helpful discussions.


\vspace{-0.5cm}

\begin{thebibliography}{99}


\bibitem{Gubser:2008px}
  S.~S.~Gubser,
  Phys.\ Rev.\  D {\bf 78} (2008) 065034
  [arXiv:0801.2977 [hep-th]].


\bibitem{H31}
  S.~A.~Hartnoll, C.~P.~Herzog and G.~T.~Horowitz,
  Phys.\ Rev.\ Lett.\  {\bf 101} (2008) 031601
  [arXiv:0803.3295 [hep-th]].


\bibitem{H32}
  S.~A.~Hartnoll, C.~P.~Herzog and G.~T.~Horowitz,
  JHEP {\bf 0812}, 015 (2008)
  [arXiv:0810.1563 [hep-th]].




\bibitem{dh}
  F.~Denef and S.~A.~Hartnoll,
  arXiv:0901.1160 [hep-th].


\bibitem{Gauntlett:2009zw}
  J.~P.~Gauntlett, S.~Kim, O.~Varela and D.~Waldram,
  JHEP {\bf 0904} (2009) 102
  [arXiv:0901.0676 [hep-th]].





\bibitem{Duff:1984sv}
  M.~J.~Duff, B.~E.~W.~Nilsson and C.~N.~Pope,
  Phys.\ Lett.\  B {\bf 139} (1984) 154.

\bibitem{Gubser:2009qm}
  S.~S.~Gubser, C.~P.~Herzog, S.~S.~Pufu and T.~Tesileanu,
  arXiv:0907.3510 [hep-th].
  
  
\bibitem{Franco:2009yz}
     S.~Franco, A.~Garcia-Garcia and D.~Rodriguez-Gomez,
arXiv:0906.1214 [hep-th] .

\bibitem{pw}
  C.~N.~Pope and N.~P.~Warner,
  Phys.\ Lett.\  B {\bf 150} (1985) 352;
  C.~N.~Pope and N.~P.~Warner,
  Class.\ Quant.\ Grav.\  {\bf 2} (1985) L1.


\bibitem{romans}
  L.~J.~Romans,
  Phys.\ Lett.\  B {\bf 153} (1985) 392.

\bibitem{Gubser:2008wz}
  S.~S.~Gubser and F.~D.~Rocha,
  Phys.\ Rev.\ Lett.\  {\bf 102} (2009) 061601
  [arXiv:0807.1737 [hep-th]].
  

\end{thebibliography}
\end{document}